\documentstyle[12pt]{article}

\input{tcilatex}

\begin{document}

\title{W state generation and effect of cavity photons on the purification
of dot-like single quantum well excitons}
\author{C.-M. Li$^{1}$, Y.-N. Chen$^{1}$, C.-W. Luo$^{1}$, J.-Y. Hsieh$^{2}$%
, D.-S. Chuu$^{1\thanks{%
Corresponding author email address: dschuu@mail.nctu.edu.tw;
Fax:886-3-5725230; Tel:886-3-5712121-56105.}}$ \\
$^{1}$Institute and Department of Electrophysics, National Chiao\\
Tung University, Hsinchu 30050, Taiwan.\\
$^{2}$Department of Mechanical Engineering, Ming Hsin University \\
of Science and Technology, Hsinchu 30401,Taiwan.}
\maketitle

\begin{abstract}
A scheme of three-particle entanglement purification is presented in this
work. The physical system undertaken for investigation is dot-like single
quantum well excitons independently coupled through a single microcavity
mode. The theoretical framework for the proposed scheme is based on the
quantum jump approach for analyzing the progress of the trible-exciton
entanglement as a series of conditional measurement has been taken on the
cavity field state. We first investigate how cavity photon affects the
purity of the double-exciton state and the purification efficiency in
two-particle protocol. Then we extend the two-particle case and conclude
that the three-exciton state can be purified into W state, which involves
the one-photon-trapping phenomenon, with a high yield. Finally, an
achievable setup for purification using only modest and presently feasible
technologies is also proposed.
\end{abstract}

Quantum information and computation bring together the concepts from
classical information science, computation theory, and quantum physics. The
EPR-Bell\cite{1} correlation, which occurs only when the quantum systems are 
{\it entangled}, generates the modern division between quantum and classical
information theory. The regulation methods of quantum information processing
such as quantum teleportation\cite{2}, quantum data compression\cite{3,4},
and quantum cryptography\cite{5} rely on the transmission of maximally
entangled pairs over quantum channels between a sender and a receiver. As it
is well known, the quantum channels are always noisy due to inevitable
interactions with environments\cite{6}. The pairs shared by the parties may
become undesired mixed states. For this reason, great attentions have been
focused on the agreement of entanglement purification\cite{7}, schemes of
entanglement distillation\cite{8}, and the decoherence mechanisms of quantum
bits (qubits) in a reservoir\cite{9}.

Although the environment leads to the decoherence of the qubits, it may play
an active role on the formation of the nonlocal effect under well
considerations. Recently, many investigations\cite{10} have been devoted to
the considerations of the reservoir-induced entanglement between two remote
qubits. By utilizing the the coupled field of bipartite systems\cite%
{11,12,13} or, generally speaking, manipulating a third system which
interacts with two remote qubits\cite{14}, many schemes have been proposed
to enhance the entanglement fidelity. The scenario usually associates with a
composite quantum system consisting two two-level subsystems inside a
leaking optical cavity with the purpose of taking unintermitted monitor on
the coupled field state. These proposals are only applicable in specific
situations that the subsystems and the coupled field are in well prepared
states, or in certain instant that the degree of entanglement is maximum.
However, more general situations are still lacking in a entanglement
distillation or entanglement generation process, especially the extension of
the proposed two-particle schemes based on conditioned measurements to the
protocol of multi-particle entanglement generation. This issue is crucial
for both the real application of quantum communication\cite{15} and the
study on the mechanism of multi-particle entanglement.

In this paper, we present a notion of three-particle entanglement
purification and an achievable setup using only modest and presently
feasible technologies. The physical system undertaken for entanglement
purification is dot-like single quantum well exciton coupled through a
single microcavity mode. The proposed scheme is depicted in Fig.1. The whole
procedure in the scheme is based on optical initialization, manipulation,
and read-out of exciton state. In it, the qubit is coded in the presence of
an exciton in a quantum well (QW), i.e. the exciton state in $i$th QW, $%
\left| e,h\right\rangle _{i}$, is considered as the logical state $\left|
0\right\rangle _{i}$, and the vacuum state, $\left| 0,0\right\rangle _{i}$,
which represents the state with no electron and hole, is coded as the
logical state $\left| 1\right\rangle _{i}$. The theoretical framework for
the proposed scheme is based on the quantum jump approach\cite{16}. To
analyze the evolution of the double-exciton entanglement, a series of
conditional measurement has been taken on the cavity field state. For
accomplishing conditional measurements faithfully, the injection and leak of
the cavity photon are controlled by means of the electro-optic effect.
Therefore, the photon field plays an active role in the procedure of
purification. According to our results of analysis, the proposed scheme can
search for the decoherence-free state, i.e. the maximally entangled state,
of the qubit-reservoir system in a time shorter than the characteristic time
of the qubit system. Furthermore, a high purification yield and nearly unit
of the entanglement fidelity can be acquired in our proposal.

First, we consider that our system consists of only two dot-like single QWs
embedded inside a single-mode microcavity with the same coupling constant.
We assume that the lateral size of the two QWs are sufficiently larger than
the Bohr radius of excitons but smaller than the wavelength of the photon
fields. Therefore, the dipole-dipole interactions and other nonlinear
interactions can be neglected. The cavity mode is assumed to be resonant
with the exitons. Under the rotating wave approximation, the unitary time
evolution of the system is then governed by the interaction picture
Hamiltonian

\begin{equation}
H_{2(I)}=\sum_{n=1}^{2}\hbar \gamma (a\sigma _{n}^{+}+a^{+}\sigma _{n}^{-}),
\end{equation}%
where $\gamma $ is the coupling constant, $a^{+}$ ($a$) is the creation
(annihilation) operator of the cavity field, and $\sigma _{n}^{+}$ ($\sigma
_{n}$) represents the creation (annihilation) operator of the excitons in
the $i$th QW.

The dynamics of the double-dot excitons in the cavity consists of three
processes: one- and two-photon absorption, one- and two-photon emission, and
no photon absorption or emission. Since the formation of the maximally
entangled exciton is a photon-trapping phenomenon, the main idea of the
proposed scheme is to find out an eigenstate involved in the process of
one-photon absorption and emission. Obviously, if the total number of quanta
in the system is, $m$, there would be an eignstates of the Hamiltonian
associates with the photon-trapping process, namely%
\begin{equation}
\left| \phi \right\rangle =\frac{1}{\sqrt{2}}(\left| 1\right\rangle
_{1}\otimes \left| 0\right\rangle _{2}-\left| 0\right\rangle _{1}\otimes
\left| 1\right\rangle _{2})\otimes \left| m-1\right\rangle _{c},
\end{equation}%
where $\left| m-1\right\rangle _{c}$ refers to the cavity field state with $%
m-1$ quanta. Once the system is in this state, the whole system does not
decay at all. Accordingly, keeping the cavity mode in state $\left|
m-1\right\rangle _{c}$ , i.e. destroying the population of the cavity field
state, paves the way to generate the entangled excitons

\begin{equation}
\left| \psi \right\rangle =\frac{1}{\sqrt{2}}(\left| 1\right\rangle
_{1}\otimes \left| 0\right\rangle _{2}-\left| 0\right\rangle _{1}\otimes
\left| 1\right\rangle _{2}).
\end{equation}%
Therefore, taking a measurement on the cavity mode in order to see whether
its state remains in the state or not is the key in our scheme.

For two-particle entanglement purification, the two dot-like single QWs and
cavity mode is prepared in the vacuum state, $\left| \phi _{0}\right\rangle
=\left| 0\right\rangle _{1}\otimes \left| 0\right\rangle _{2}\otimes \left|
0\right\rangle _{c}$. A laser pulse, which is tuned to the lowest interband
excitation energy, is then applied on the first QW. It can promote an
electron from the valence-band to the conduction-band, hence the state, $%
\left| \phi _{0}\right\rangle =\left| 1\right\rangle _{1}\left|
0_{2}\right\rangle \left| 0\right\rangle _{c}$, prepared for purification is
obtained. For the sake of generality and purpose of distillation, the state
of QW can be any mixed state, $\rho _{\psi }$, except the vacuum state.
Next, a pulse with $m$ photons is injected into the microcavity. For the
feasibility and the modest technology requirements, we consider here the
injection of single photon, i.e. $m=1.$ The total number of quantum count of
the system is two.

As the single-photon has been injected into the cavity, the total system
will evolve with time, and is governed by the operator

\begin{equation}
U(t)=\left( 
\begin{array}{cccc}
2\gamma ^{2}a(G-K)a^{+}+1 & -i\gamma aS & -i\gamma aS & 2\gamma ^{2}a(G-K)a
\\ 
-i\gamma Sa^{+} & \frac{1}{2}(\cos (\mu t)+1) & \frac{1}{2}(\cos (\mu t)-1)
& -i\gamma Sa \\ 
-i\gamma Sa^{+} & \frac{1}{2}(\cos (\mu t)-1) & \frac{1}{2}(\cos (\mu t)+1)
& -i\gamma Sa \\ 
2\gamma ^{2}a^{+}(G-K)a^{+} & -i\gamma a^{+}S & -i\gamma a^{+}S & 2\gamma
^{2}a^{+}(G-K)a+1%
\end{array}%
\right) ,
\end{equation}%
where $\mu ^{2}=K^{-1}=2\gamma ^{2}(2a^{+}a+1)$, $G=K\cos (\mu t)$, and $%
S=\mu ^{-1}\sin (\mu t)$. If the system evolves without interruption, it
will go into a QW1- QW2-cavity field entangled state. If we take a
measurement on the cavity field state at some instant, the number of the
photon count of the detector may be one, two, or zero. Since the
single-photon state $\left| 1\right\rangle _{c}$ involves the
photon-trapping phenomenon, we can infer that the double-QW will evolve into
a maximal entangled state if the cavity mode stays in state $\left|
1\right\rangle _{c}$ via the quantum jump approach\cite{16}. After measuring
the cavity field state, injecting a subsequent photon is necessary for the
sake of keeping the photon in its state. We then let the whole system evolve
for another period of time $\tau $. Again, we proceed to measure the cavity
photon to make sure whether its quantum is one or not. If the cavity photon
remains in single-photon, the repetition continues; if not, the whole
procedure should be started over. Therefore, after several times of
successful repetitions, the state of double-exciton progresses into the state

\begin{equation}
\rho _{\psi _{N}}=(_{c}\left\langle 1\right| U(\tau )\left| 1\right\rangle
_{c})^{N}\rho _{\psi _{i}}(_{c}\left\langle 1\right| U(-\tau )\left|
1\right\rangle _{c})^{N}/P_{N},
\end{equation}%
where

\begin{equation}
P_{N}=\text{Tr}[(_{c}\left\langle 1\right| U(\tau )\left| 1\right\rangle
_{c})^{N}\rho _{\psi _{i}}(_{c}\left\langle 1\right| U(-\tau )\left|
1\right\rangle _{c})^{N}],
\end{equation}%
is the probability of success for measuring a single-photon after $N$ times
of repetitions. The superoperator, $_{c}\left\langle 1\right| U(\tau )\left|
1\right\rangle _{c}$, reveals the significant fact that it will evolve to
the projection operator, $\left| \psi \right\rangle \left\langle \psi
\right| $, as the successful repetitions increases. This result comes from
the fact that the superoperator, $_{c}\left\langle 1\right| U(\tau )\left|
1\right\rangle _{c}$, has only one eigenvalue which its absolute value
equals to one\cite{14}, here, the corresponding eigenvector is just the
photon-trapping state. Thus the double-exciton state will become a maximal
entangled state as long as the repetitions is sufficient large. By Eq. (6),
the probability of success can be evaluated

\begin{equation}
P_{N,1}=\frac{1}{2}(1+\cos (\sqrt{6}\gamma \tau )^{2N}),
\end{equation}%
and the fidelity is

\begin{eqnarray}
F_{N,1} &=&\left\langle \psi \right| \rho _{\psi _{N}}\left| \psi
\right\rangle  \nonumber \\
&=&\frac{1}{2\cos (\sqrt{6}\gamma \tau )^{2N}}.
\end{eqnarray}%
Here, if the product, $\gamma \tau $, is set to be $n\pi /\sqrt{6}$, $%
n=1,2,...,$ then the fidelity of the double-exciton state will approach to
one and the probability of success will be $1/2$ as $N$ increases.

If the state of the cavity field is kept in $n-$photon state, rather than
the state with single photon, Eq. (7) and Eq. (8) can be generalized to

\begin{equation}
P_{N,n}=\frac{1}{2}(1+\cos (\sqrt{2(n+1)}\gamma \tau )^{2N}),
\end{equation}%
and

\begin{equation}
F_{N,n}=\frac{1}{2\cos (\sqrt{2(n+1)}\gamma \tau )^{2N}}.
\end{equation}%
It reveals that the number of measured cavity photon and evolution period
play important roles in the trade-off between $P_{N,n}$ and $F_{N,n}$. We
can choose a set of $(n,\gamma \tau )$ such that the fidelity progresses to
one at the least repetitions, however, in the same time it causes the
probability to reduce to a minimum. This is the usual case for purification.
On the other hand, one can also find a suitable such that the probability
goes to one. In this case, the system will not evolve with time and is
similar to the Zeno paradox with {\em finite} duration between two
measurements.

The concept of two-particle entanglement purification discussed above can be
generalized to three-particle case directly. Here, the same approximations
have been made as in the double-exciton case, and the formulation of
three-exciton dynamics can be derived straightforwardly via the interaction
Hamiltonian

\begin{equation}
H_{3(I)}=\hbar \gamma \left( 
\begin{array}{cccccccc}
0 & a & a & 0 & a & 0 & 0 & 0 \\ 
a^{+} & 0 & 0 & a & 0 & a & 0 & 0 \\ 
a^{+} & 0 & 0 & a & 0 & 0 & a & 0 \\ 
0 & a^{+} & a^{+} & 0 & 0 & 0 & 0 & a \\ 
a^{+} & 0 & 0 & 0 & 0 & a & a & 0 \\ 
0 & a^{+} & 0 & 0 & a^{+} & 0 & 0 & a \\ 
0 & 0 & a^{+} & 0 & a^{+} & 0 & 0 & a \\ 
0 & 0 & 0 & a^{+} & 0 & a^{+} & a^{+} & 0%
\end{array}%
\right) .
\end{equation}%
The superoperator that governs the progress of the three dot-like single QWs
with single-photon can be worked out:

\begin{eqnarray}
_{c}\left\langle 1\right| e^{iH_{3(I)}\tau }\left| 1\right\rangle _{c}
&=&g\left| g\right\rangle \left\langle g\right| +W_{1}\left|
W_{1}\right\rangle \left\langle W_{1}\right| +T_{1}\left| T_{1}\right\rangle
\left\langle T_{1}\right| ++T_{2}\left| T_{2}\right\rangle \left\langle
T_{2}\right|  \nonumber \\
&&+e\left| e\right\rangle \left\langle e\right| +W_{2}\left|
W_{2}\right\rangle \left\langle W_{2}\right| +T_{3}\left| T_{3}\right\rangle
\left\langle T_{3}\right| ++T_{4}\left| T_{4}\right\rangle \left\langle
T_{4}\right| ,
\end{eqnarray}%
where $g$, $e$, $W_{1}$, $W_{2}$, $T_{1}$, $T_{2}$, $T_{3}$ and $T_{4}$ are
functions of $\tau $ which correspond to the orthonormal eigenvectors

\begin{eqnarray}
\left| g\right\rangle &=&\left| 0\right\rangle _{1}\otimes \left|
0\right\rangle _{2}\otimes \left| 0\right\rangle _{3},  \nonumber \\
\left| W_{1}\right\rangle &=&\frac{1}{\sqrt{3}}(\left| 1\right\rangle
_{1}\otimes \left| 0\right\rangle _{2}\otimes \left| 0\right\rangle
_{3}+\left| 0\right\rangle _{1}\otimes \left| 1\right\rangle _{2}\otimes
\left| 0\right\rangle _{3}+\left| 0\right\rangle _{1}\otimes \left|
0\right\rangle _{2},  \nonumber \\
\left| T_{1}\right\rangle &=&\frac{1}{\sqrt{2}}(\left| 1\right\rangle
_{1}\otimes \left| 0\right\rangle _{2}\otimes \left| 0\right\rangle
_{3}-\left| 0\right\rangle _{1}\otimes \left| 0\right\rangle _{2}\otimes
\left| 1\right\rangle _{3}),  \nonumber \\
\left| T_{2}\right\rangle &=&\frac{1}{\sqrt{6}}(\left| 1\right\rangle
_{1}\otimes \left| 0\right\rangle _{2}\otimes \left| 0\right\rangle
_{3}-2\left| 1\right\rangle _{1}\otimes \left| 0\right\rangle _{2}\otimes
\left| 1\right\rangle _{3}+\left| 0\right\rangle _{1}\otimes \left|
0\right\rangle _{2}\otimes \left| 1\right\rangle _{3}),  \nonumber \\
\left| e\right\rangle &=&\left| 1\right\rangle _{1}\otimes \left|
1\right\rangle _{2}\otimes \left| 1\right\rangle _{3},  \nonumber \\
\left| W_{2}\right\rangle &=&\frac{1}{\sqrt{3}}(\left| 0\right\rangle
_{1}\otimes \left| 1\right\rangle _{2}\otimes \left| 1\right\rangle
_{3}+\left| 1\right\rangle _{1}\otimes \left| 0\right\rangle _{2}\otimes
\left| 1\right\rangle _{3}+\left| 1\right\rangle _{1}\otimes \left|
1\right\rangle _{2}\otimes \left| 0\right\rangle _{3}),  \nonumber \\
\left| T_{3}\right\rangle &=&\frac{1}{\sqrt{2}}(\left| 0\right\rangle
_{1}\otimes \left| 1\right\rangle _{2}\otimes \left| 1\right\rangle
_{3}-\left| 1\right\rangle _{1}\otimes \left| 1\right\rangle _{2}\otimes
\left| 0\right\rangle _{3}),  \nonumber \\
\text{and }\left| T_{4}\right\rangle &=&\frac{1}{\sqrt{6}}(\left|
0\right\rangle _{1}\otimes \left| 1\right\rangle _{2}\otimes \left|
1\right\rangle _{3}-2\left| 0\right\rangle _{1}\otimes \left| 1\right\rangle
_{2}\otimes \left| 0\right\rangle _{3}+\left| 1\right\rangle _{1}\otimes
\left| 1\right\rangle _{2}\otimes \left| 0\right\rangle _{3}).
\end{eqnarray}%
Here $T_{1}=T_{2}$ and $T_{3}=T_{4}$ are two two-fold degenerate eigenvalues
of the superoperator $_{c}\left\langle 1\right| e^{iH_{3(I)}\tau }\left|
1\right\rangle _{c}$.

If the initial state is set equal to 
\begin{equation}
\left| \phi _{i}\right\rangle =\left| 1\right\rangle _{1}\otimes \left|
0\right\rangle _{2}\otimes \left| 0\right\rangle _{3}\otimes \left|
0\right\rangle _{c},
\end{equation}%
, the probability of success for finding the exciton state is in state $%
\left| W_{1}\right\rangle $ can then be worked out analytically:%
\begin{equation}
P_{N}=\frac{1}{3}(\cos (\sqrt{10}\gamma \tau )^{2N}+2\cos (\gamma \tau
)^{2N}),
\end{equation}%
and the corresponding fidelity of the three-exciton state is

\begin{eqnarray}
F_{N} &=&\left\langle W_{1}\right| \rho _{\psi _{N}}\left| W_{1}\right\rangle
\nonumber \\
&=&\frac{\cos (\sqrt{10}\gamma \tau )^{2N}}{\cos (\sqrt{10}\gamma \tau
)^{2N}+2\cos (\gamma \tau )^{2N}}.
\end{eqnarray}%
Here, we can set $\gamma \tau $ to be $n\pi /\sqrt{10}$, $n=1,2,...$, then
the fidelity of the three-exciton state approaches to unit as $N$ increases;
meanwhile the probability of success is $1/3$. A special case, $\gamma \tau
=\pi /\sqrt{6}$, is demonstrated in Fig. 2. Fig. 2(a) describes the
trade-off between $P_{N}$ and $F_{N}$. The purification yield $%
Y_{N}=\prod_{i=0}^{N}P_{i}$, which measures the fraction of surviving pairs,
is also presented in Fig. 2(b).

Greenberger-Horn-Zeilinger (GHZ)\cite{17} and W states\cite{18,19} are two
kinds of three-particle maximally entanglement. The former involves
three-photon trapping phenomenon which can be described through the state
vector

\begin{equation}
\left| \psi _{\text{GHZ}}\right\rangle =\frac{1}{\sqrt{3}}(\left|
0\right\rangle _{1}\otimes \left| 0\right\rangle _{2}\otimes \left|
0\right\rangle _{3}-\left| 1\right\rangle _{1}\otimes \left| 1\right\rangle
_{2}\otimes \left| 1\right\rangle _{3}).
\end{equation}%
and W state is a one-photon trapping state

\begin{equation}
\left| \psi _{\text{W}}\right\rangle =\frac{1}{\sqrt{3}}(\left|
1\right\rangle _{1}\otimes \left| 0\right\rangle _{2}\otimes \left|
0\right\rangle _{3}+\left| 0\right\rangle _{1}\otimes \left| 1\right\rangle
_{2}\otimes \left| 0\right\rangle _{3}+\left| 0\right\rangle _{1}\otimes
\left| 0\right\rangle _{2}\otimes \left| 1\right\rangle _{3}).
\end{equation}%
Although these states associated with different physical phenomenon, they
possess the same degree of entanglement\cite{18,19}. In our scheme, any
initial mixed state of the three-exciton state can be purified into W state
except the vacuum. However, the symmetry properties of qubit-environment
interactions for GHZ state is quite different from three-exciton W state
except the vacuum, thus it can not be generated via the approach of
conditional measurement.

The whole concepts for the experiments are schematically shown in Fig. 3. A
Ti:sapphire laser supplies the pulse light source for all of the operation
photons in our devices such as resonant photon, $2\omega $ photon, and pulse 
${\bf E}$ field. The timing between all of the operation photons could be
controlled precisely by delay stages. Also, the detection of photodiodes
could be precisely triggered by laser pulses. Firstly, one of the dot-like
single quantum wells is excited by the $3eV$ photon from $2\omega $
generator performed through a nonlinear crystal, e.g. BBO or LBO crystal.
Then, a resonant photon with vertical linear polarization generated via a
quartz plate is injected into the cavity which is constructed by the ZnTe
with both Au films. Meanwhile, through the pulse ${\bf E}$ field\cite{15}
with appropriate magnitude\cite{18} and pulse width of $\symbol{126}3.3ps$
as shown in Fig. 4, the linear polarization of injected photon is rotated
from vertical to horizontal via the Electro-optic effect in ZnTe\cite{19}.
After the sufficient evolution time ($T_{ev}$ $\symbol{126}20ps$, the time
period, $\tau $, between two measurements) with dot-like single quantum well
excitons, the photon in cavity can be leaked out the cavity by a pulse ${\bf %
E}$ field with suitable timing and detected by a single photon avalanche
diode (SPAD, detector 2 in Fig. 3). This procedure would be repeated until
finishing the purification. During this procedure, the photoluminescence
lifetime of dot-like single quantum well exciton can be measured by the
other SPAD as shown in Fig. 4 (detector 1 in Fig. 3).

To summarize, a three-particle entanglement purification scheme based on
conditioned measurements has been proposed in this work. We investigate the
entanglement generation of dot-like single quantum well excitons coupled
through a single microcavity mode. As shown in Eq. (9) and Eq. (10), we
first consider how the cavity photon affects the purity of the exciton state
and the purification efficiency in two-particle protocol. The trade-off
between $P_{N,n}$ and $F_{N,n}$ reveals whether the double excitons can be
purified into a photon trapping, decoherence-free state efficiently
depending on the number of photon counts in the repeated measurements.
Moreover, we conclude that the three-exciton state can be purified into W
state with a well yield but smaller than that in two-particle case. Finally,
a feasible experimental setup for optical initialization, manipulation, and
read-out of exciton state is also presented.

\bigskip This work is supported partially by the National Science Council,
Taiwan under the grant number NSC 92-2120-M-009-010.

\bigskip

{\Huge Figure Caption}\bigskip

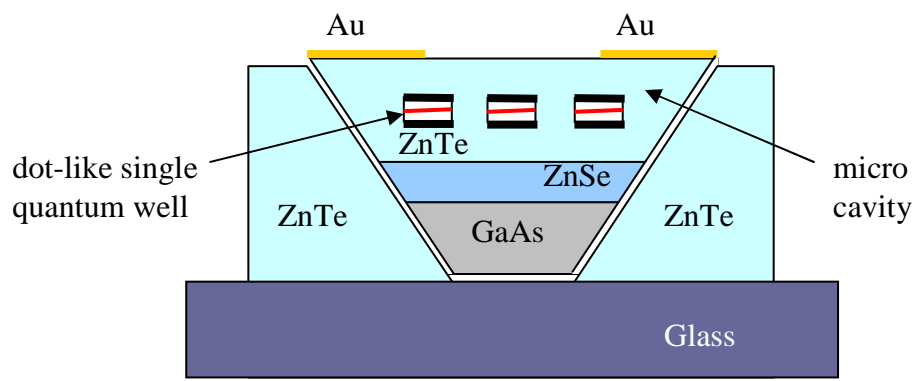
Fig. 1. The quantum devices with triple dot-like quantum wells inlaid in a
microcavity which is constructed by a ZnTe medium and two Au mirrors, will
be prepared by the MBE, the e-beam lithography, and the conventional
semiconductor processing.

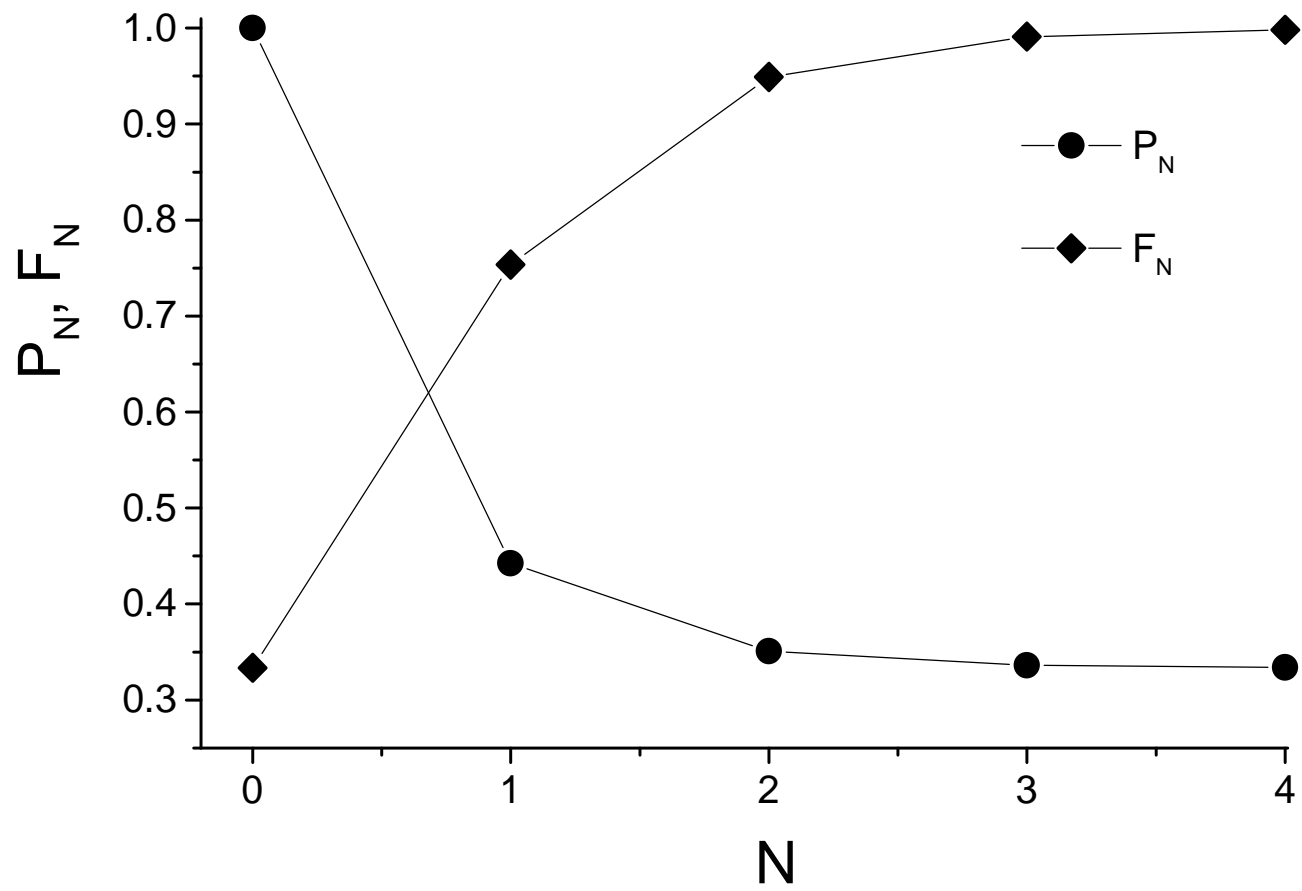
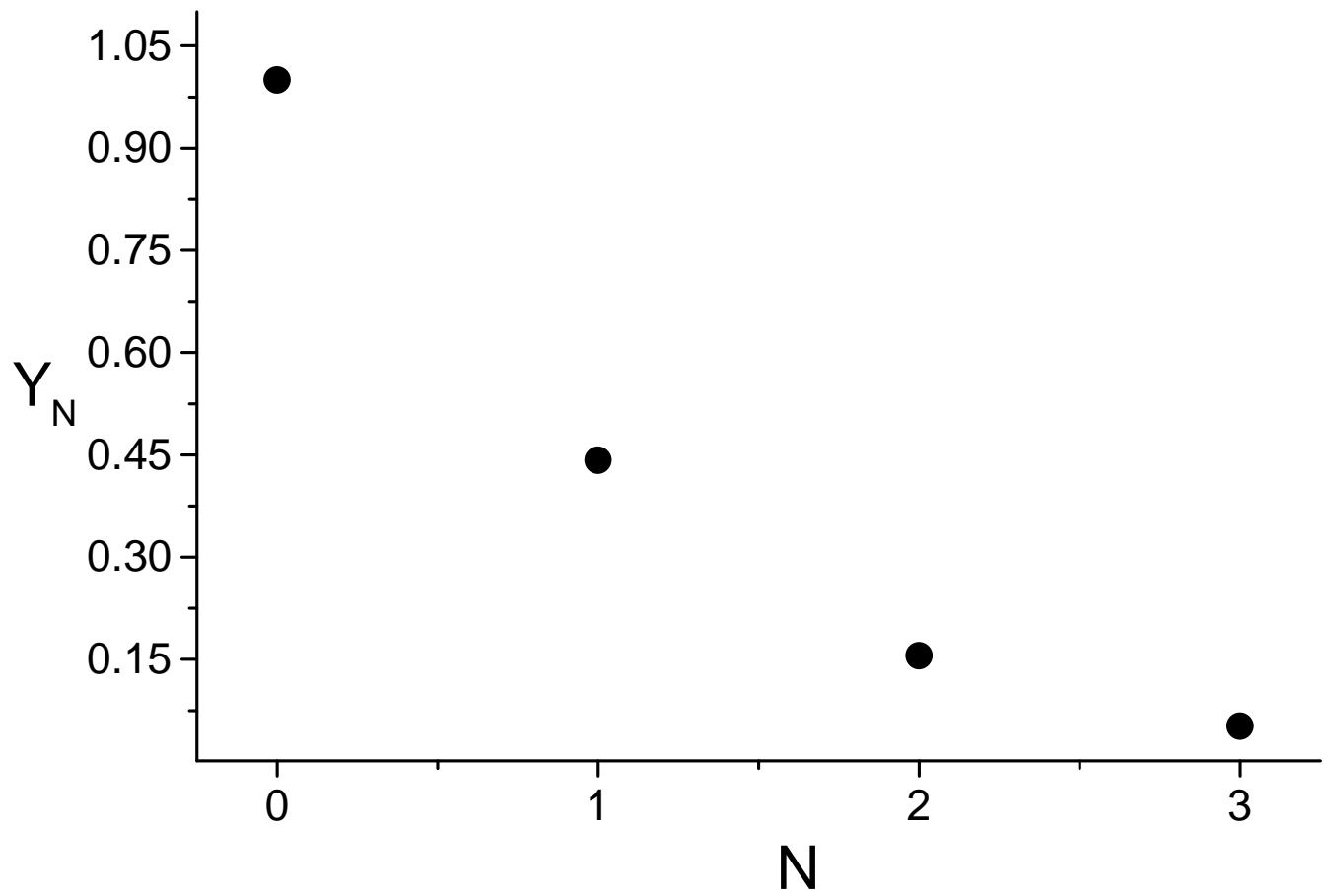
Fig. 2. (a) The probability of success for measuring the same photon state, $%
P_{N}$, and the fidelity of the exciton state, $F_{N}$, and (b) the
corresponding yield, $Y_{N},$ for W state generation.

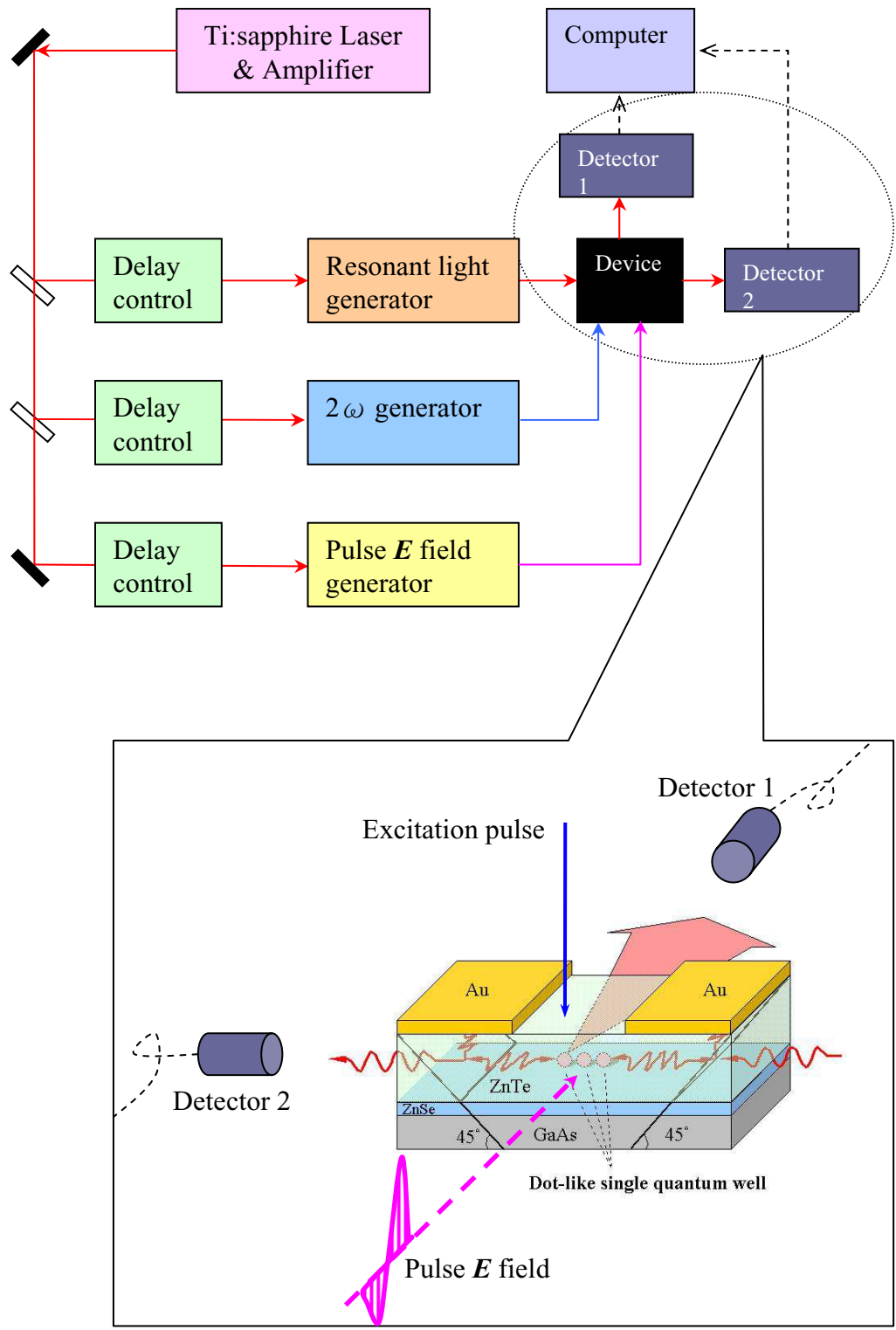
Fig. 3. The whole experimental concepts for optical initialization,
manipulation, and read-out of exciton state.

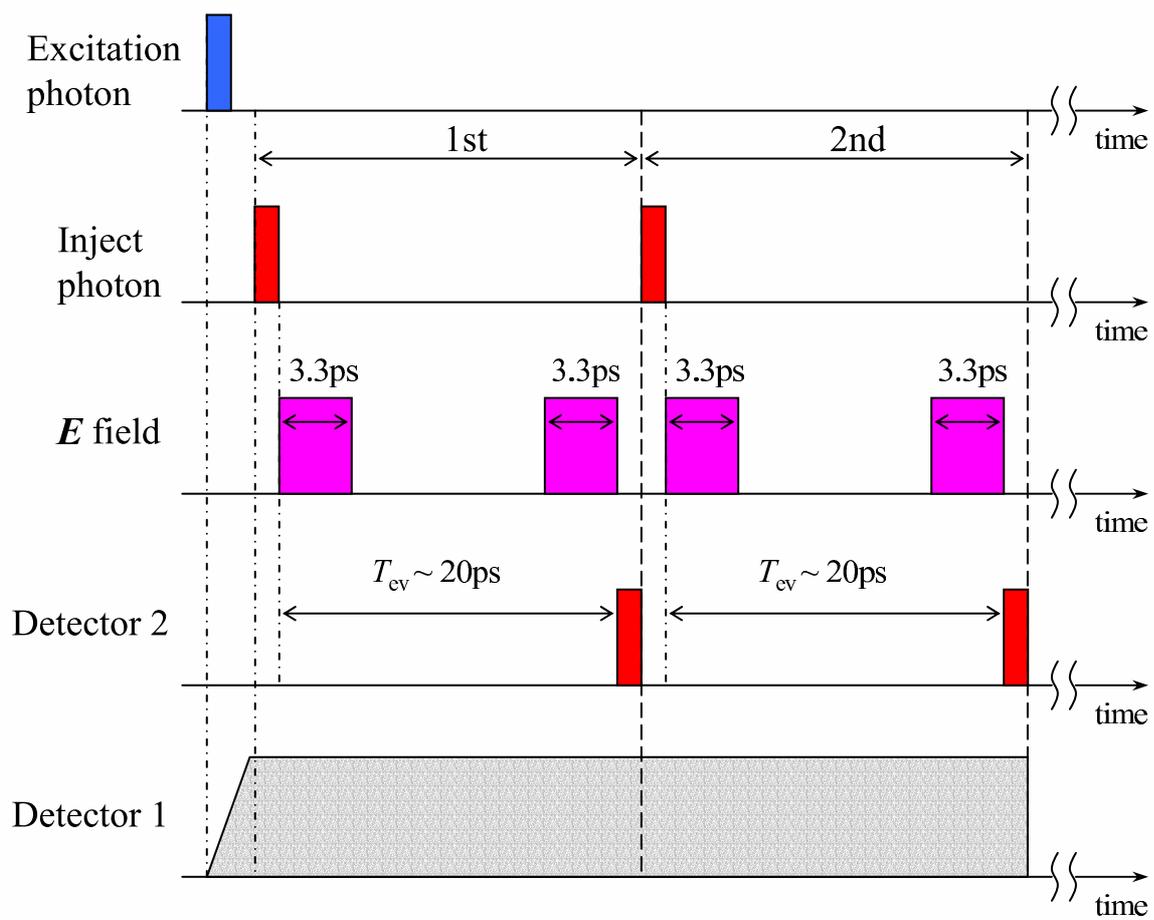
Fig. 4. The flowchart for W state generation. The inset quantum device is
the same with the device in Fig. 1.


\begin{thebibliography}{99}
\bibitem{20} Actually, the pulse E field is the ``THz radiation'' which
could be easily generated from GaAs semiconductor or several nonlinear
crystal such as GaSe [21, 22].
\end{thebibliography}
\end{document}